# ON TSALLIES NONEQUILIBRIUM ENTROPY EVOLUTION

XING XIU-SAN

( Department of Physics, Beijing Institute of Technology, Beijing 100081, China )

(email:xingxiusan@gmail.com)

**Abstract**    In this paper we derived a 6N dimensional nonhomogeneous evolution equation of Tsallis nonequilibrium entropy ; presented a formula for entropy production rate (i.e. the law of entropy increase) for Tsallis entropy only when its index $q > 0$, otherwise the law of entropy increase does not hold when $q < 0$ or $q = 0$.

## 1. Introduction

Entropy is one of the most important concept and physical quantity in physics. The entropy change represents the evolution direction of macroscopic nonequilibrium physical systems. Although the entropy and the law of entropy increase have been extensively studied, a lot of major physical properties of nonequilibrium entropy have not been fully understood so far. From the viewpoint of microscopic statistical theory, in all the field of entropy, since for a long time past, we only have a Boltzmann's entropy formula $S = k \ln W$ which gives microscopic statistical explanation of entropy, but did not have any statistical equation and formula to describe the evolution of nonequilibrium entropy. This just remains to be explored and solved. More than ten years ago, we first obtained a nonlinear evolution equation of Boltzmann-Gibbs(BG) nonequlibrium entropy and a formula for entropy production rate in 6N and 6 dimendional phase space. In recent twenty-five years C.Tsallies proposed a nonextensive entropy for generalizing BG-entropy. Hence he and his collaborators published many papers and two books. Now, the questions rise: Does Tsallies nonequilibrium entropy (i.e. nonextensive nonequilibrium entropy) also obey any evolution equation? What is the form of this equation if it exists? More importantly, does it also hold the law of entropy increase? If yes, general or partial? In this paper we shall answer these questions.

## 2. The Liouville diffusion equation

Considering that the Liouville equation is the time-reversed symmetrical, it is cannot be to deduce and explain the irreversibility of nonequilibrium macroscopic systems, the law of entropy increase and the time-reversed asymmetrical hydrodynamic equations etc in a rigorous and unified fashion, more than fifteen years ago, the author proposed the Liouville diffusion equation in 6N dimensional phase space[1-4]

$$\frac{\partial \rho}{\partial t} = [H, \rho] + D\nabla_q^2 \rho = -\dot{X} \cdot \nabla_X \rho + D\nabla_q^2 \rho \tag{1}$$

as a new fundamental equation of nonequilibrium statistical physics in place of the Liouville equation.

Where

$$\dot{X} = (\nabla_p H, -\nabla_q H), \quad -\nabla_X \cdot (\dot{X}\rho) = -\dot{X} \cdot \nabla_X \rho = [H, \rho] \tag{2}$$

$\rho = \rho(X,t) = \rho(q,p,t) = \rho(q_1, q_2, \cdots, q_N; p_1, p_2, \cdots, p_N; t)$ is the ensemble probability density.

$H = H(X) = H(q, p) = H(q_1, q_2, \ldots q_N; p_1, p_2, \ldots p_N)$ is Hamiltonian of the system, $X = (q, p)$ is the state vector in 6N dimensional phase space, $q$ and $p$ are the set vectors $q = (q_1, q_2, \ldots q_N)$ and $p = (p_1, p_2, \ldots p_N)$; $q_i$ and $p_i$ are the generalized coordinates and momenta of the ith particles respectively. $D$ is the diffusion coefficient of the particles. Compared with the Liouville equation, this equation has an extra stochastic diffusive term in coordinate subspace, which represents that the particles in statistical thermodynamic systems not only drifts in phase space but also diffuses in coordinate subspace. Hence it is microscopic time-reversed asymmetrical and reflects the irreversibility of statistical thermodynamic processes.

For the equilibrium states, the ensemble probability density $\rho_0 = \rho_0(X)$ satisfies

$$\frac{\partial \rho_0}{\partial t} = [H, \rho_0] + D\nabla_q^2 \rho_0 = -\dot{X} \cdot \nabla_X \rho_0 + D\nabla_q^2 \rho_0 = 0 \tag{3}$$

We shall derive nonequilibrium entropy evolution equation and the formula for entropy production rate for Tsallis entropy by means of Eqs (1) (3) in the following sections.

## 3. Nonequilibrium entropy evolution equation

In order to comparing with BG entropy's results, we firstly cite them here, then derive Tsallies entropy's results.

BG nonequilibrium entropy in 6N-dimensional phase space can be defined as[2-6]

$$S_G(t) = -k\int \rho(X, t)\ln \frac{\rho(X, t)}{\rho_0(X)} d\Gamma + S_{G0}$$

$$= \int S_X d\Gamma + S_{G0} \tag{4}$$

where $k$ is the Boltzmann constant,
BG entropy density in 6N dimensional phase space

$$S_X = -k\rho \ln \frac{\rho}{\rho_0} \tag{5}$$

or 
$$S_G(t) = -k\int \rho(X, t)\ln\rho(X, t)d\Gamma = \int S_X d\Gamma \tag{4a}$$

$$S_X = -k\rho(X, t)\ln\rho(X, t) \tag{5a}$$

In this paper, we use formulas (4) rather than (4a) as the definition of nonequilibrium entropy, the reason will be explained by the following formulas (15a).

Taking derivative of both sides of formula (4) with respect to time $t$ and using the Liouville diffusion Eqs. (1) and (3), we obtain BG 6N dimensional nonequilibrium entropy evolution equation.[2-4]

$$\frac{\partial S_X}{\partial t} = -\nabla_X \cdot (\dot{X}S_X) + D\nabla_q^2 S_X$$

$$+ \frac{D}{k\rho}[(\nabla_q \ln\rho)S_X - \nabla_q S_X]^2 \tag{6}$$

Where $-\nabla_X \cdot (\dot{X}S_X) = [H, S_X]$

The entropy production density[1-4]

$$\sigma_G = kD\rho(\nabla_q \ln\frac{\rho}{\rho_0})^2 = \frac{D}{k\rho}[(\nabla_q \ln\rho)S_X - \nabla_q S_X]^2 \qquad (7)$$

Eq(6) shows that the time rate of change of nonequilibrium entropy density (the term on the left-hand side) is caused together by its drift (the first term on the right-hand side), diffusion (the second term on the right-hand side) and production (the third term on the right-hand side). This means that entropy being an important extensive physical quantity, its density distribution in nonequilibrium statistical thermodynamic systems is always nonuniform, nonequilibrium and changes with time and space.

According to Tsallis work[7-9], Tsallis nonequilibrium entropy in 6N dimensional phase space can be defined as

$$S_q(t) = k\frac{1 - \int [\rho(\mathbf{x},t)]^q d\Gamma}{q-1} = k\frac{\int \rho(\mathbf{x},t)d\Gamma - \int [\rho(\mathbf{x},t)]^q d\Gamma}{q-1}$$
$$= \int S_{q\mathbf{x}} d\Gamma \qquad (8)$$

Tsallis entropy density in 6N dimensional phase space

$$S_{q\mathbf{x}}(t) = k\frac{\rho - \rho^q}{q-1} \qquad (9)$$

$$\lim_{q \to 1} S_q(t) = k\frac{1 - \int \rho[1 + (q-1)\ln\rho]d\Gamma}{q-1}$$
$$= -k\int \rho \ln\rho d\Gamma \qquad [\rho = \rho(\mathbf{x},t)] \qquad (8a)$$

$$\lim_{q \to 1} S_{q\mathbf{x}}(t) = k\frac{\rho - \rho[(1+(q-1))\ln\rho]}{q-1} = -k\rho\ln\rho \qquad (9a)$$

In order to corresponding to formula (4), Tsallis entropy should be defined as

$$S_q(t) = k\frac{1 - \int \rho_0(\mathbf{x})\left[\rho(\mathbf{x},t)/\rho_0(\mathbf{x})\right]^q d\Gamma}{q-1} + S_m$$
$$= k\frac{\int \rho(\mathbf{x},t)d\Gamma - \int \rho_0(\mathbf{x})\left[\rho(\mathbf{x},t)/\rho_0(\mathbf{x})\right]^q d\Gamma}{q-1} + S_m$$
$$= \int S_{q\mathbf{x}} d\Gamma + S_m \qquad (10)$$

Tsallis entropy density

$$S_{q\mathbf{x}}(t) = k\frac{\rho - \rho_0(\rho/\rho_0)^q}{q-1} \tag{11}$$

$$\lim_{q \to 1} S_q(t) = k\frac{1 - \int \rho_0 \left\{ \frac{\rho}{\rho_0}\left[1 + (q-1)\ln\frac{\rho}{\rho_0}\right] \right\} d\Gamma}{q-1} + S_m$$

$$= -k\int \rho \ln\frac{\rho}{\rho_0} d\Gamma + S_m \tag{10a}$$

$$\lim_{q \to 1} S_{q\mathbf{x}}(t) = k\frac{\rho - \rho_0\left\{\frac{\rho}{\rho_0}\left[1 + (q-1)\ln\frac{\rho}{\rho_0}\right]\right\}}{q-1} = -k\rho \ln\frac{\rho}{\rho_0} \tag{11a}$$

Taking derivative of both sides of formula (11) with respect to time and using the Liouville diffusion Eqs(1) and (3), we obtain the time rate of change of Tsallis entropy density in 6N dimensional phase space.

$$\frac{\partial S_{q\mathbf{x}}}{\partial t} = \frac{k}{q-1}\frac{\partial}{\partial t}\left(\rho - \rho_0^{1-q}\rho^q\right)$$

$$= \frac{k}{q-1}\left[\frac{\partial \rho}{\partial t} - q\rho_0^{1-q}\rho^{q-1}\frac{\partial \rho}{\partial t} - (1-q)\rho_0^{-q}\rho^q\frac{\partial \rho_0}{\partial t}\right]$$

$$= \frac{k}{q-1}\left\{\left(1 - \rho_0^{1-q}\rho^{q-1}\right)\left[-\nabla_X \cdot \left[(\dot{X} - D\nabla_q \ln\rho)\rho\right]\right]\right.$$

$$+ (1-q)\rho_0^{1-q}\rho^{q-1}\left[-\nabla_X \cdot \left[(\dot{X} - D\nabla_q \ln\rho)\rho\right]\right]$$

$$\left. - (1-q)\rho_0^{-q}\rho^q\left[-\nabla_X \cdot \left[(\dot{X} - D\nabla_q \ln\rho_0)\rho_0\right]\right]\right\}$$

$$= \frac{k}{q-1}\left\{-\left(1 - \rho_0^{1-q}\rho^{q-1}\right)\left[\nabla_X \cdot \left[(\dot{X} - D\nabla_q \ln\rho)\rho\right]\right]\right.$$

$$- \left[\nabla_X\left(1 - \rho_0^{1-q}\rho^{q-1}\right)\right]\left[(\dot{X} - D\nabla_q \ln\rho)\rho\right]$$

$$\left. + (1-q)D\rho_0^{1-q}\rho^q\nabla_q^2 \ln\frac{\rho}{\rho_0}\right.$$

$$+ (1-q) D \rho_0^{1-q} \rho^q \left( \nabla_q \ln \frac{\rho}{\rho_0} \right) \nabla_q \ln \rho_0 \}$$

$$= -\nabla_x \cdot \left[ S_{q\mathbf{x}} (\dot{X} - D \nabla_q \ln \rho) \right] - kD \left( \frac{\rho}{\rho_0} \right)^q \nabla_q \cdot \left( \rho_0 \nabla_q \ln \frac{\rho}{\rho_0} \right)$$

$$= -\nabla_x \cdot (\dot{X} S_{q\mathbf{x}}) + D \nabla_q^2 S_{q\mathbf{x}} + kD \nabla_q \cdot \left[ \rho_0 \left( \frac{\rho}{\rho_0} \right)^q \nabla_q \ln \frac{\rho}{\rho_0} \right]$$

$$- kD \left( \frac{\rho}{\rho_0} \right)^q \nabla_q \cdot \left( \rho_0 \nabla \ln \frac{\rho}{\rho_0} \right) \tag{12}$$

In obtaining Eq（12）we substitute formula (11) into the right-hand side of Eq(12) many times.

Since

$$k D \nabla_q \left[ \rho (\rho/\rho_0)^q \nabla_q \ln \frac{\rho}{\rho_0} \right] - kD (\rho/\rho_0)^q \nabla_q \left( \rho_0 \nabla_q \ln \frac{\rho}{\rho_0} \right)$$

$$= kD \left[ \nabla_q (\rho/\rho_0)^q \rho_0 \nabla_q \ln \frac{\rho}{\rho_0} \right]$$

$$= kq D \rho_0^{1-q} \rho^q \left( \nabla_q \ln \frac{\rho}{\rho_0} \right)^2 \tag{13}$$

Therefore, we have

$$\frac{\partial S_{q\mathbf{x}}}{\partial t} = -\nabla_\mathbf{x} \cdot (\dot{X} S_{q\mathbf{x}}) + D \nabla_q^2 S_{q\mathbf{x}} + kqD\rho_0^{1-q}\rho^q \left( \nabla_q \ln \frac{\rho}{\rho_0} \right)^2 \tag{14}$$

This is Tsallis 6N dimensional nonequilibrium entropy evolution equation. However, there is no $S_{q\mathbf{x}}$ in its third term on the right hand side and we cannot make $S_{q\mathbf{x}}$ appear in it too. Hence it is a nonhomogeneous equation.

We note that $q$ is different from $q$ in Eqs. (14). $q$ is a index in Tsallis entropy, but $\mathbf{q}=(\mathbf{q}_1, \mathbf{q}_2, \cdots \mathbf{q}_N)$ in $\nabla_q$ are the set vectors.

Eq (14) is Tsallis nonequilibrium entropy evolution equation in 6N dimensional phase space, which is presented here firstly. It also shows that the time rate of change of Tsallis nonequilibrium entropy density is caused together by drift, diffusion and production ($q>0$) or disappearance ($q<0$).and or zero ($q=0$).

Comparing Eq (14) with Eq (6), it is easy to see that the first terms and the second terms on the right-hand sides of Eq (14) and Eq (6) are the same in mathematical form and physical meaning, but the third terms are different.

## 4. Formula for entropy production rate

According to expressions (7), BG entropy production rate in 6N and 6 dimensional phase space are

$$P_G = \frac{\partial_i S_G}{\partial t} = \int \sigma_G d\Gamma = kD \int \rho (\nabla_q \ln \frac{\rho}{\rho_0})^2 d\Gamma \qquad (15)$$

Being similar to the definition of strain or elongation percentage[10] $\varepsilon = \ln(l/l_0)$ of solid materials under the action of complex stress, we can define a new physical parameter of nonequilibrium system, that is the departure percentage from equilibrium of the ensemble probability density of nonequilibrium system in $6N$ dimensional phase space as

$$\theta = \ln \frac{\rho}{\rho_0} \qquad (16)$$

Using the number density of micro-states $\omega_0$ and $\omega$ and the total number of micro-states $W_0$ and $W$ for equilibrium and nonequilibrium states, which satisfy the relations $\omega_0 = W_0 \rho_0$ and $\omega = W\rho$ respectively, then (16) changes to

$$\theta = \ln \frac{\omega}{\omega_0} - \ln \frac{W}{W_0} \approx \frac{\Delta \omega}{\omega_0} - \frac{\Delta W}{W_0} \qquad (17)$$

Substituting (17) into (15), then BG entropy production rate in 6N dimensional phase space changes to[4]

$$P_G = \frac{\partial_i S_G}{\partial t} = kD \int \rho (\nabla_q \ln \frac{\omega}{\omega_0})^2 d\Gamma = kD \int \rho (\nabla_q \theta)^2 d\Gamma = kD \overline{(\nabla_q \theta)^2}$$

That is $\qquad P_G = \frac{\partial_i S_G}{\partial t} = kD \overline{(\nabla_q \theta)^2} \geq 0 \qquad (18)$

Where $\overline{(\nabla_q \theta)^2} = \int \rho (\nabla_q \theta)^2 d\Gamma$ is the average value of the square of the space gradient of the departure percentage from equilibrium.

It should be pointed out here that entropy increase $P=0$ at equilibrium state are direct conclusions of formulas (18), it need not any additive condition. If expression (4a) is applied as the definition of nonequilibrium entropy, although the mathematical form of nonequilibrium entropy evolution equation (6) does not change, the third terms on the right-hand side of equation, i.e. the expressions of the entropy increase rate changes into

$$P_G' = \int \sigma_G d\Gamma = kD \int \rho (\nabla_q \ln \rho)^2 d\Gamma \geq 0 \qquad (15a)$$

Here if we want that $P'=0$ at equilibrium state, it need $\nabla_q \rho = 0$ and $\nabla_q f = 0$. In general, this is not reasonable. That is the reason why expression (4) but not expression (4a) is applied as the definition of the nonequilibrium entropy.

Formulas (18) is the concise statistical formulas for entropy production rate in 6N and 6 dimensional phase space. It is also the quantitative concise statistical formula for the law of entropy increase for BG entropy.

According to the third term on the right-hand side in Tsallis nonequilibrium entropy evolution equation (14) the Tsallis entropy production rate in 6N dimensional phase space is

$$P_q = \frac{\partial_i S_q}{\partial t} = kqD \int \rho (\rho/\rho_0)^{q-1} (\nabla_q \ln \frac{\rho}{\rho_0})^2 d\Gamma \qquad (19)$$

Substituting (16) into (19), then the Tsallis entropy production rate changes to

$$P_q = \frac{\partial_i S_q}{\partial t} = kqD \int \rho [e^{(q-1)\theta}](\nabla_q \theta)^2 d\Gamma = kqD \overline{[e^{(q-1)\theta}](\nabla_q \theta)^2} \qquad (20)$$

Note, when $q>0$, $P_q = =\frac{\partial_i S_q}{\partial t} \geq 0$; when $q<0$, $P_q = =\frac{\partial_i S_q}{\partial t} < 0$ and $q=0$, $P_q = =\frac{\partial_i S_q}{\partial t} = 0$.

That is, when index $q>0$, formula (20) is the concise statististical formula for the Tsallis entropy production rate ,i.e. the concise statistical formula for the law of entropy increase of Tsallis entropy. However, when index $q<0$ or $q=0$, the entropy production is negative or equal to zero. In these two cases, the Tsallis entropy does not obey the law of entropy increase.

Formulas (18)(20) clearly tell us that the entropy production rate of a nonequilibrium physical system is only determined by two physical parameters: diffusion coefficient $D$ and the departure percentage from equilibrium $\theta$ (not including the known constant $k$). It can be used in calculating the entropy production rate of the actual physical topics when $D$ and $\theta$ are known. The diffusion coefficient $D$ can be calculated and measured from experiment. The departure percentage from equilibrium $\theta$, a new defined physical parameter, its physical meaning is clear, whose introduction not only makes the formula for the entropy production rate simple and clear, but also makes that it can play the role of an physical parameter to describe quantitatively how far a nonequilibrium system is from equilibrium as if the strain describes the deformation of the solids.

It can be seen that when index $q=1$, $S_q=S_G$, then Eq (14)=Eq (6), formula (20)=formula (18).

## 5. Conclusions

Nonequilibrium entropy evolution equations describe the evolution law of nonequilibrium entropy and the evolution process of nonequilibrium systems. They play a central role in nonequilibrium entropy theory. For BG entropy, they show that the time rate of change of nonequilibrium entropy density is caused together by its drift,diffusion and production. However,for Tsallis entropy it shows that that the time rate of change of nonequilibrium entropy density is caused together by its drift,diffusion and production(q>0) or disappearance(q<0) and or zero(q=0).

The law of entropy increase is a fundamental law in nature. For BG entropy,the formula for entropy production rate, i.e. the formula for the law of entropy increase demonstrates that the macroscopic entropy production of a nonequilibrium physical system is caused by spatially stochastic and inhomogeneous departure from equilibrium of the number density of micro-states. For Tsallis entropy, the formula for entropy production rate, i.e. the formula for the law of entropy increase is presented only when its index $q>0$. Its physical meaning is the same as BG entropy's, but its expression is more complicate. However, the law of entropy increase does not hold when $q<0$ or $q=0$.

When index $q=1$, Tsallis entropy changes into BG entropy, both Tsallis nonequilibrium entropy evolution equation and the formula for the law of entropy increase also change into BG entropy's.